\documentclass[preprint,3p,times]{elsarticle}
\journal{Separation and Purification Technology | Accepted on June 16, 2024}

\usepackage[version=3]{mhchem} 
\usepackage{siunitx}
\usepackage{tabularx}
\usepackage{caption}
\usepackage{multicol}
\usepackage{amsmath}
\usepackage{amssymb}
\usepackage{multirow}
\usepackage{xcolor}
\usepackage{hyperref}
\hypersetup{colorlinks=true, linkcolor=blue}

\DeclareSIUnit\bar{bar}
\author[inst1]{Subhadeep Dasgupta}
\author[inst1]{Amal R S}
\author[inst1]{Prabal K. Maiti\corref{pkm}}

\title
  {Unifying Mixed Gas Adsorption in Molecular Sieve Membranes  and MOFs using Machine Learning}

\affiliation[inst1]{
        organization={Department of Physics, Indian Institute of Science},
        city={Bangalore},
        postcode={560012}, 
        state={Karnataka},
        country={India}
        }
\cortext[pkm]{
	Corresponding author,
	\ead{maiti@iisc.ac.in}
	Tel: +918022932865
}

\begin{document}

\begin{frontmatter}

\begin{abstract}
  Recent machine learning models to accurately obtain gas adsorption isotherms focus on polymers or metal-organic frameworks (MOFs) separately.
  The difficulty in creating a unified model that can predict the adsorption trends in both types of adsorbents is challenging, owing to the diversity in their chemical structures.
  Moreover, models trained only on single gas adsorption data are incapable of predicting adsorption isotherms for binary gas mixtures.
  In this work, we address these problems using feature vectors comprising only the physical properties of the gas mixtures and adsorbents.
  Our model is trained on adsorption isotherms of both single and binary mixed gases inside carbon molecular sieving membrane (CMSM), together with data available from CoRE MOF database.
  The trained models are capable of accurately predicting the adsorption trends in both classes of materials, for both pure and binary components.
  ML architecture designed for one class of material, is not suitable for predicting the other class, even after proper training, signifying that the model must be trained jointly for proper predictions and transferability.
  The model is used to predict with good accuracy the \ce{CO2} uptake inside CALF-20 framework.
  This work opens up a new avenue for predicting complex adsorption processes for gas mixtures in a wide range of materials.
\end{abstract}

\end{frontmatter}
\section{Introduction}
In recent times, the rapid design and manufacture of nanoporous materials has sparked interest in the chemical industry as a means to replace existing solvent-based filtration with highly selective adsorbent frameworks.
Adsorption-based filtration are positioned to provide higher yield, while also consuming less energy during chemical separation processes~\cite{ma2020manufacturing}.
Unlike other extraction techniques, nanofiltration operates under moderate temperature and pressure conditions, and increases safety during operations~\cite{sholl2016seven,kumar201750th}.
Nanoporous materials include a diverse range of materials, all having completely different chemical structures leading to a wide range of physicochemical properties.
The most common categories include metal-organic frameworks (MOFs)~\cite{furukawa2013chemistry}, porous organic cages (POCs)~\cite{tozawa2009porous}, covalent-organic frameworks (COFs)~\cite{cote2005porous}, and zeolites~\cite{gottardi2012natural, kianfar2022recent}.
There has also been significant developments in understanding formation of nanoporous membranes, making them another viable choice for chemical filtration.
Advances have been made in the fabrication of nanochannels~\cite{mao2020designing}, polymers~\cite{robeson1999polymer}, and mixed matrix membranes~\cite{kamble2021review}.
MOFs are currently the most studied class of porous crystalline materials, made up of repeating units of small building units consisting of metal sites connected to organic molecules.
Polymeric membranes are fabricated from growing connections of individual monomers.
The connectivity of these monomers, their growth rate, cross-linking, and chain length can greatly influence their structural properties and adsorption trends~\cite{maiti2004structure, song2022scalable, dasgupta2022influence, maity2023efficient,canivet2014water,lopez2024water,carta2013efficient}.
Research is also being carried out to fabricate MOF-based membranes for enhanced gas separations~\cite{qian2020mof}.
The different stoichiometry involved during fabrication gives rise to a diverse variety of membranes, each showing unique physical and chemical properties.
The vast zoo of materials allows us to find the perfect candidate for a particular filtration task, on the basis of its affinity to accept one type of gas penetrant from different gas mixtures.

The adsorption performance of complex materials can be evaluated using computational techniques, allowing pre-screening of materials prior to manufacture, greatly cutting production costs for industries.
Molecular dynamics (MD) simulations, help in understanding the formation of the adsorbent material.
Grand canonical Monte Carlo (GCMC) simulations of the adsorbent kept in contact with a virtual reservoir of gas molecules provide accurate prediction of the membrane gas uptake capacity.
However, the bottleneck of these simulations arise from their increasing system sizes.
Performing simulations (or experiments) to assess the performance on all possible combinations of adsorbents and gas mixtures is thus still a challenge.
To overcome the computational limitations, predictive algorithms powered by machine learning (ML) models, trained on available data, is currently being explored to predict gas separation of newly discovered materials~\cite{anderson2020adsorption, tang2018efficiently}.
However, the vast majority of existing literature data primarily focus on single component adsorption, thereby raising concerns on a trained model's ability to make accurate predictions for gas mixtures~\cite{barnett2020designing}.
In current times there has been a peak in industrial and academic interests of using ML algorithms towards materials discovery and their prediction capacities~\cite{butler2018machine, chen2020critical, jackson2019recent, wang2023machine}. 
Since the output of all such models are correlated with the input dataset, training on MOFs alone restricts predictions on polymer membranes and vice versa.
Data driven techniques, such as transfer learning~\cite{bozinovski2020reminder}, fine-tuning~\cite{liu2022few}, and predicting with large language models (LLMs)~\cite{hoffmann2022training} are not applicable in such cases since the nature of adsorption is inherently different in different classes of materials~\cite{al2020guidelines}.
Additionally, there the available database for membranes is not as robust as for MOFs, making it difficult to validate the predictions.
The type of input features used for MOFs (fingerprints) during training may not be available for other nanomaterials, making it impossible for an existing model to be applied on polymer membranes.

ML assisted workflows have been exceptionally powerful in performing high throughput screening of millions of possible MOF for finding optimum thermal and mechanical stability and also for targeted applications~\cite{chong2020applications, shi2020machine, huang2023machine, bag2020machine}.
Energy based descriptors have been used to rapidly model and understand MOFs with exceptional  \ce{H2} storage capacities~\cite{bucior2019energy, bobbitt2019molecular, ahmed2019exceptional}.
Likewise, similar efforts have resulted in identification of composite MOFs for use in \ce{CO2} storage in harsh environmental setting of low to high pressures~\cite{fan2024unconventinoal,magnin2023abnormal, zhang2022machine, burner2020high, guda2020machine,choudhary2022graph}.
The methodology has also proven to be useful for prediction gas adsorption in general in MOFs~\cite{hung2022chemistry, fernandez2013atomic, altintas2021machine}.
The role of publicly available large databases has also facilitated the study of nanopores in the structures which further enriches the understanding of gas adsorption processes and aid in future designing of MOFs~\cite{luo2022mof, bag2021interaction, krishnapriyan2021machine, korolev2020transferable,datar2020beyond, anderson2018role,jablonka2020big}.
Sophisticated state-of-the-art tools like genetic algorithms, transfer learning and tailored ML models have also been proposed for guided and intelligent choice of MOFs towards a variety of use cases including \ce{CH4} sensing~\cite{gustafson2019intelligent}, accurate MOF forcefield development~\cite{vandenhaute2023machine} and gas separations~\cite{hu2022machine, moghadam2019structure, zhang2019machine,cooper2023metal}.
The success of the above techniques in predicting material properties and molecular interactions have also been utilized to study gas adsorption in polymeric membranes~\cite{gormley2021machine, kunalan2021efficient}.
ML has also empowered predicting the gas diffusivities and permeabilities of a wide range of materials~\cite{barnett2020designing, shastry2024machine,hasnaoui2017neural} which otherwise requires performing careful  simulations, particularly for gas mixtures~\cite{monteleone2022advanced, dasgupta2023trajectory, neyertz2010trajectory, fraga2018novel}.
Techniques have also been developed to account for incomplete datasets during the training process of an ML model~\cite{yuan2021imputation}, obtaining fractional free volume of membranes~\cite{tao2023machine}, and predicting a given polymer membrane's adsorption selectivity and their performance towards gas separation~\cite{osman2024machine,veliouglu2024predictive,gao2023understanding,zhao2023improved, abdollahi2023simulating, pan2022analysis,pilz2023utilizing}.

Regardless of the vast number of advances, all  ML algorithms are restricted by their training data, limiting their scope of applicability.
In this work, we propose a solution addressing the transferability concerns of trained ML model, by training on combined data of carbon molecular sieving membranes (CMSMs) and MOFs.
The input and output features allow for both single and binary gas mixtures while also taking into consideration the presence of moisture.
The input features used are purely physical properties of a material which can be easily obtained for any class of material.
The methodology section describes the database and techniques used to train the ML models, followed by the results section which discusses our findings.
In the conclusion section, we discuss the implications and scope of this work.

\section{Method}

\subsection{Data preparation}

\begin{figure*}[ht!]
    \centering
    \includegraphics[width=0.95\linewidth]{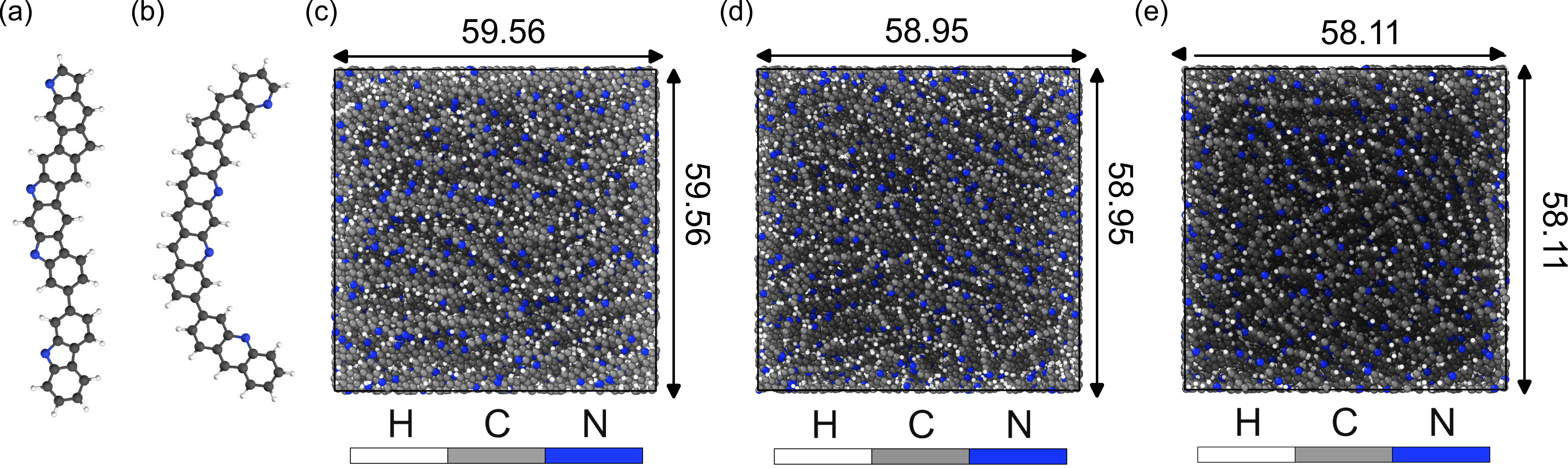}
    \caption{
        Monomers (a) pyrrole and (b) pyridine used to build different $6$F-CMSM polymers.
        Equilibrated morphologies of three representative membranes.
        The inverse density $(1/ \rho)$ of these membranes are (c) highest $(\SI{0.86}{\cubic\centi\meter\per\gram})$, (d) average $(\SI{0.79}{\cubic\centi\meter\per\gram})$, and (e) lowest $(\SI{0.69}{\cubic\centi\meter\per\gram})$.
        The simulation box lengths are in $\AA$ units.
        The bottom color legends show the constituent elements in the polymer membrane.
    }
    \label{fig:structure_CMS}

    \includegraphics[width=0.85\linewidth]{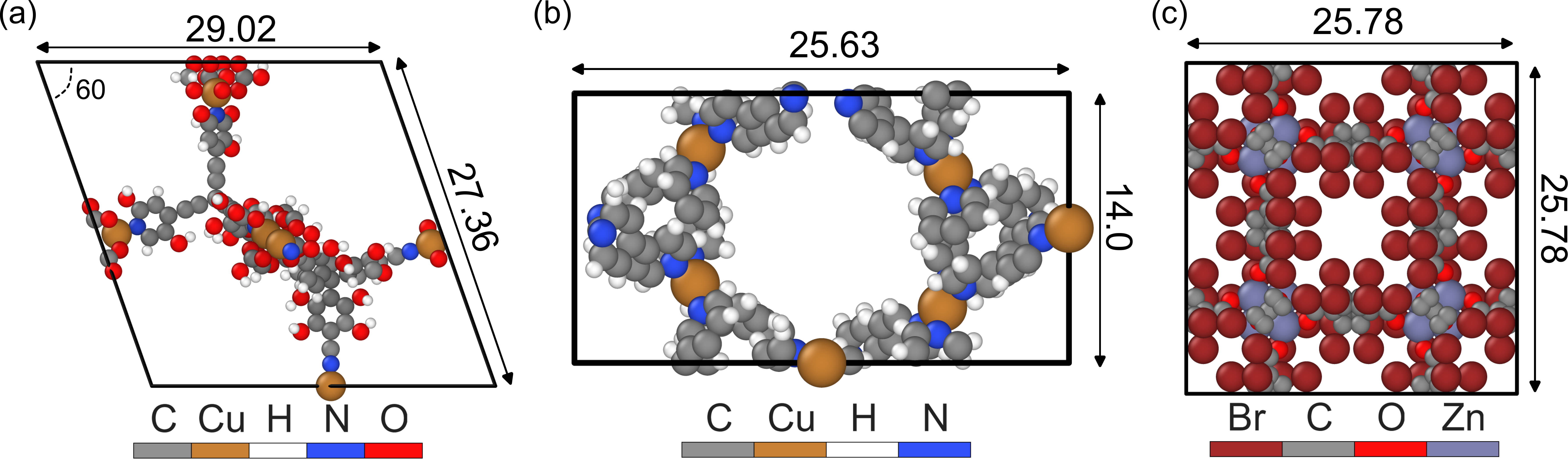}
    \caption{
        Unit cell structures of three MOFs in the dataset.
        The inverse density $(1/ \rho)$ values are (a) highest $(\SI{6.41}{\cubic\centi\meter\per\gram})$, (b) average $(\SI{1.22}{\cubic\centi\meter\per\gram})$, and (c) lowest $(\SI{0.46}{\cubic\centi\meter\per\gram})$.
        The unit cell lengths of the MOFs are in $\AA$ units.
        The bottom color legends show the constituent elements in the MOFs.
    }
    \label{fig:structure_MOF}
\end{figure*}

To obtain gas adsorption data for polymers we require datasets comprising significant variations in the membrane's structural properties.
In our previous work~\cite{dasgupta2022influence}, we have shown that the physical properties of a carbon molecular sieving membrane derived from 6FDA/BPDA-DAM polymer precursor ~\cite{kumar2019highly} ($6\text{F-CMSM}$) is highly tunable, based on the chain length of individual polymer strands.
The 6F-CMSMs are built using pyrrole and pyridine monomers (Figure~\ref{fig:structure_CMS}a,~\ref{fig:structure_CMS}b), distributed randomly throughout each polymer chain.
Variation in porosity of equilibrated membranes also explain the different gas adsorption performances observed in experiments owing to pyrolisis of the precursor at different temperatures.
The bonded and non-bonded interactions are computed using parameters of the modified Dreiding forcefield~\cite{Mayo1990dreiding, Roy2020investigations}.
The MD simulations to obtain the density equilibrated 6F-CMSM structures were performed using compression-decompression algorithms using LAMMPS~\cite{thompson2022lammps} discussed in our earlier works~\cite{dasgupta2022influence, Roy2020investigations}.
The TraPPE forcefield~\cite{Trappe} is used to compute the bonded and non-bonded interactions of the gases with each other and the adsorbent framework.
Lorentz-Berthelot mixing rule is used to compute the interaction of different atom types.
For equilibration, we follow a simulation protocol comprising a series of compression-decompression algorithms in NPT ensemble for \SI{100}{\nano\second} described in detail in our previous work~\cite{dasgupta2022influence}.
The final equilibrated system corresponds to temperature $\SI{300}{\kelvin}$ and pressure of $\SI{1}{\bar}$.
The equilibrated structures of 6F-CMSM are obtained using all-atom MD simulations.
Figure~\ref{fig:structure_CMS}c to ~\ref{fig:structure_CMS}e shows three representative structures of the membrane arranged in decreasing order of inverse density $(1/\rho)$.
Different  physical properties such as gravimetric surface area (GSA), helium void fraction $(\xi)$, density $(\rho)$, largest included sphere $(D_i)$, and largest free spheres $(d_{max})$ of the different 6F-CMSM are obtained using the equilibrated MD structures using RASPA~\cite{Dubbeldam2016raspa} and Zeo++~\cite{willems2012algorithms}, enabling us to create a dataset for adsorption of different gases at different temperature $(T)$, pressure $(P)$, and vapor pressures.
These quantities are used as input features to describe the adsorbent material.
GSA, $\rho, D_i, d_{max}$ of the equilibrated $6$F-CMSMs are computed using Zeo++ ~\cite{willems2012algorithms}.
$\xi$ is obtained using Widom insertion of helium atom in the membrane\cite{Dubbeldam2016raspa}, as implemented in RASPA.

\begin{table*}[h]
    \centering
    \caption{Kinetic diameters, $d (\AA)$, of gas molecules studied in this work.}
    \label{tab:gases}
    \begin{tabular}{|c|c c c c c c c c c c c c|}
        \hline
        Gas         &
        \ce{C6H6}   &
        \ce{C2H6}   &
        \ce{C3H8}   &
        \ce{Xe}     &
        \ce{Kr}     &
        \ce{CH4}    &
        \ce{N2}     &
        \ce{O2}     &
        \ce{Ar}     &
        \ce{CO2}    &
        \ce{H2}     &
        \ce{H2O}    \\
        \hline
        $d (\AA)$   &
        $5.85 $     &
        $4.443$     &
        $4.3  $     &
        $3.96 $     &
        $3.6  $     &
        $3.758$     &
        $3.64 $     &
        $3.46 $     &
        $3.4  $     &
        $3.3  $     &
        $2.89 $     &
        $2.65 $     \\
        \hline
    \end{tabular}
\end{table*}

To obtain gas adsorption isotherms we perform a series of GCMC simulations in the $\mu$VT ensemble,  until the loading equilibrates, using RASPA~\cite{Dubbeldam2016raspa}.
The membrane is kept rigid during GCMC adsorption simulation.
Each GCMC step is composed of insertion of a gas from reservoir into the pore of $6$F-CMSM, deletion, translation, rotation, and swapping of two gas molecules, similar to the protocol used  in our previous work~\cite{dasgupta2022influence}.
The number of gases inside the membrane gradually increases starting from zero to its equilibrium value based on the environmental condition of pressure ($0-\SI{50}{bar}$) and temperature ($\SI{273}{\kelvin}$, $\SI{308}{\kelvin}$,  and $\SI{373}{\kelvin}$).
Our GCMC simulations consist of both pure and binary gas mixtures of varying mole fractions, which in turn will help our model to predict for competitive adsorption processes as well.
The kinetic diameters of the different gases are given in Table~\ref{tab:gases}.
For binary gas mixtures following compositions are studied: \ce{CO2}:\ce{CH4} are studied for 50:50, and 10:90 mole fractions, and \ce{CO2}:\ce{N2} for 20:80 mole fraction.
The additional effects of moisture on gas uptake is studied by introducing some amount of \ce{H2O} molecules in $\ce{CO2}:\ce{CH4}$ = 10:90 mixture.
Data for MOFs were taken from the Computation-Ready Experimental (CoRE) MOFs database~\cite{chung2019advances}, for which all of the above physical properties as well as gas loading capacity are available in literature~\cite{anderson2020adsorption}.
The available data is thus a subset of the complete database, containing the gases \ce{C2H6}, \ce{Xe}, \ce{Kr}, \ce{CH4}, \ce{N2}, \ce{O2}, and \ce{Ar}.
The chemical structures of three representative MOFs from this dataset are shown in Figure~\ref{fig:structure_MOF}, arranged in descending order of their inverse density.
Figure~\ref{fig:structure_CMS} and Figure~\ref{fig:structure_MOF} showcase the diversity in the free volumes inside these nanostructures, giving rise to their interesting physico-chemical properties.
From our GCMC simulations we obtain a set of 3180 adsorption data for the membrane, while there are 1741 available data for MOFs comprising all required properties.

\begin{figure}[h!]
    \centering
    \includegraphics[width=0.4\linewidth]{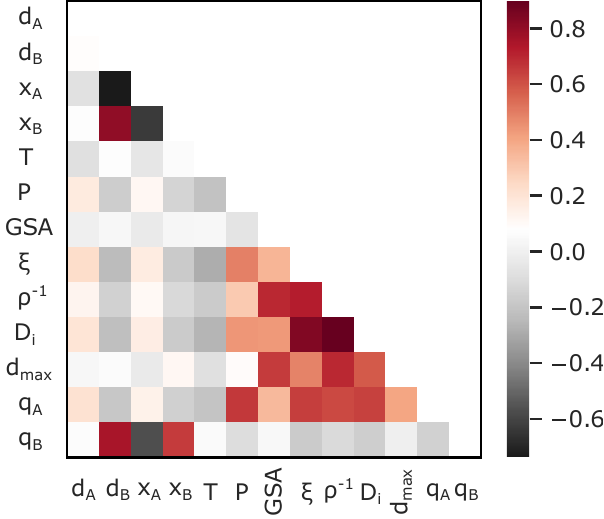}
    \caption{Correlation coefficients of input and output features.}
    \label{fig:feature_correlation}
\end{figure}

\subsection{Fingerprinting}
Each gas molecule is identified using its kinetic diameters $(d)$, followed by their mole fraction $(x)$ in the mixture.
The fingerprint used to describe each input parameter is,
\begin{align*}
    I = \big\{
        &   d_A (\AA), \enspace
            d_B (\AA), \enspace
            x_A,       \enspace
            x_B,
            \\
        &   T (K),     \enspace
            P (bar),   \enspace
            \text{GSA} (m^2g^{-1}),
            \\
        &   \xi,       \enspace
            \rho^{-1} (cm^3g^{-1}), \enspace
            D_{i} (\AA), \enspace
            d_{max} (\AA)
        \big\},
\end{align*}
where the subscripts A, B denote the two gases in a binary mixture.
The output features are the saturated gas uptake $(q)$ of the binary gas components, given by
\begin{align*}
    O = \big\{ q_{A} (mol \enspace kg^{-1}), \enspace q_{B} (mol \enspace kg^{-1}) \big\}
\end{align*}
The conditions $d_B = 0$ and $q_{B} = 0$ describe the cases for pure gas adsorption.
The correlation coefficients of the input and output features are shown in Figure~\ref{fig:feature_correlation}.
Aside from $d$ and $x$, there is no strong correlation in the parameters, making it a valid choice for fingerprinting the dataset.

\subsection{Model Training}

\begin{figure*}[t]
    \centering
    \includegraphics[width=0.9\linewidth]{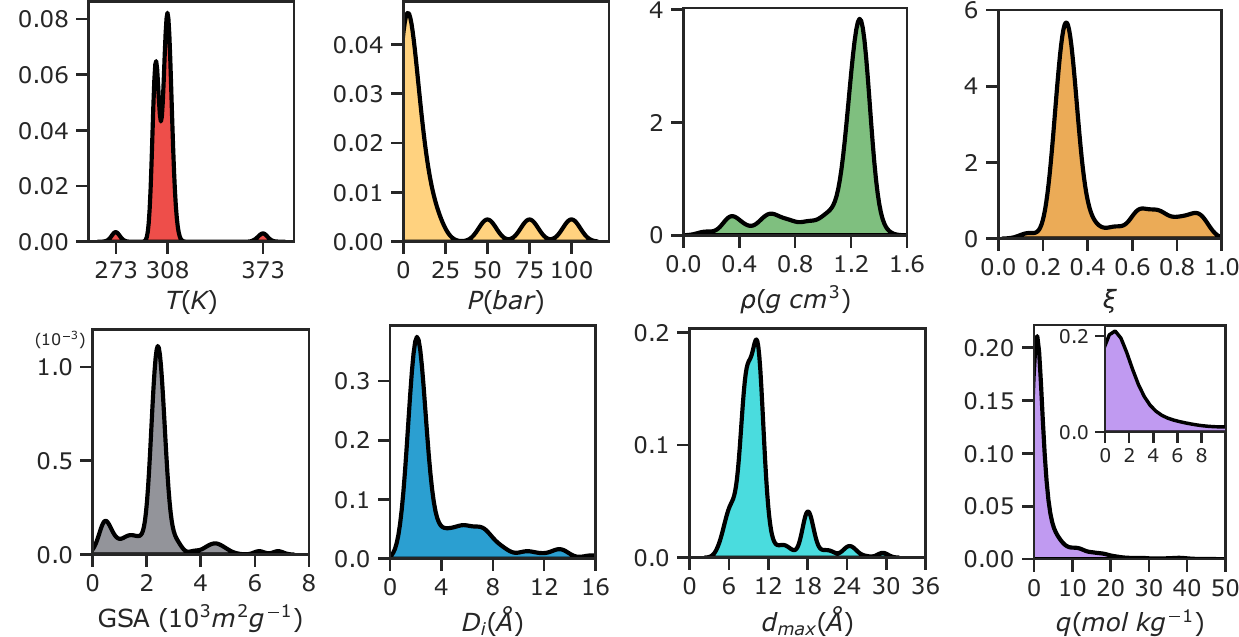}
    \caption{
    Distribution of physical properties of the combined $6$F-CMSM and MOF dataset.
    }
    \label{fig:data_distribution}
\end{figure*}

To prevent overfitting and improve the generality of predictions, a common strategy is to introduce randomness to the input feature, mimicking small noises in the dataset~\cite{bishop1995training}.
We perturb the quantities in $I(T, \dots, d_{max})$ using small random Gaussian noise.
The kinetic diameter and mole fraction of gases are excluded during this perturbation since they uniquely identify a gas molecule, and slight change in $d$ will inadvertently represent a completely different gas molecule.
Noise inherently appears in $q$ from fluctuations during the steps of GCMC simulation.
The distribution of the combined dataset is shown in Figure~\ref{fig:data_distribution}.
The combined dataset is normalized, shuffled, and split for training (80\%), validation (10\%), and testing (10\%) purposes.
The training data is fed into two ML models, artificial neural networks (ANN) in TensorFlow\cite{abadi2016tensorflow}, and XGBoost~\cite{Chen_2016}.
The ML models are kept blind to the data kept aside for testing.

Each ML model is described by a set of hyper-parameters that affect how the model learns.
NN models are made up of different nodes and layers, each node having different weights and biases to pass information to the next layer.
For NN the hyper-parameters investigated in this work include number of hidden layers $(h)$, number of nodes in each layer $(n)$, training epochs $(e)$, activation functions $(f)$, kernel initializers and regularizers, and loss functions.
We vary $n$ from $11$ to $600$.
For each $n$, we explore $h$ from $3$ to $9$.
We also explore a set of $\{n, h\}$ where $n$ decreases down from $\{2^h, \dots, 2^1\}$ for each subsequent layer.
In this case, we vary $h$ from $3$ to $16$.
Our set-up thus allows exploring shallow and deep neural networks (DNNs).
The training epochs $(e)$ are explored with and without early-stopping.
We investigate $f$ using ReLU and sigmoid functions, with orthogonal initializers, and L1, L2 regularizers.

The XGBoost model is a random forest algorithm designed to quickly navigate through the dataset, identify logical relations between input and output, thereby creating a hierarchy of hyper-branched data, terminating at different output points (leaf).
For XGBoost, we perform a thorough grid-search involving the parameters number of estimators, learning rate, depth of the tree and branches, loss reduction, and regularizers.
The performance of the models are measured using mean squared error (MSE) and mean absolute error (MAE) metrics against the testing data.
The best performing models are identified based on their training and validation losses of the output predictions w.r.t. testing data which serves as our ground truth, which serves as the reference for comparing the ML predictions.

\section{Results and discussion}

\begin{figure}[h!]
    \centering
    \includegraphics[width=0.7\linewidth]{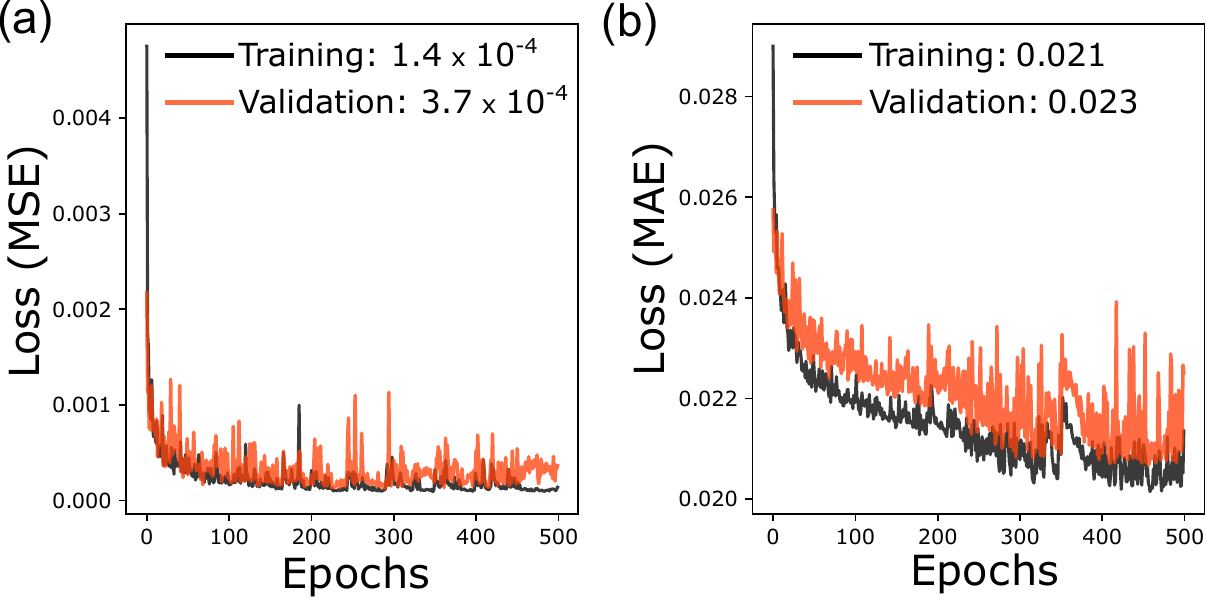}
    \caption{
    Training and validation loss vs. epochs for a neural network having $(n, h) = (160, 7)$ using (a) MSE and (b) MAE metrics respectively.
    }
    \label{fig:loss}

    \centering
    \includegraphics[width=0.7\linewidth]{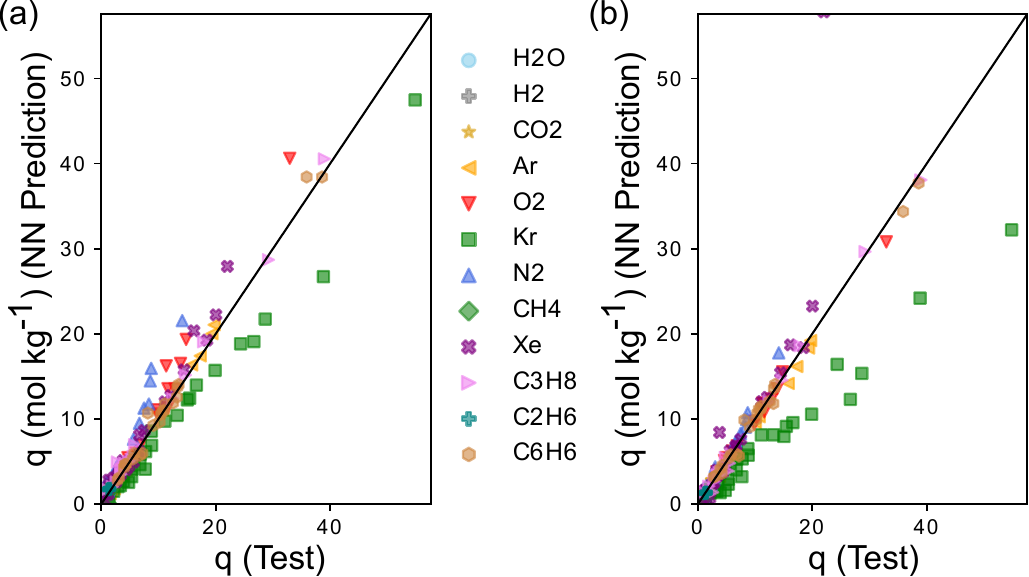}
    \caption{
    NN predictions with $(n, h) = (160, 7)$ using (a) MSE and (b) MAE metrics respectively.
    The different gas molecules are shown in different symbols shown in the legend.
    The 100\% agreement line is shown in black as a guide to the eyes.
    }
    \label{fig:agreement}
\end{figure}

We explore the various configurations of the different NN hyper-parameters to identify the optimum configuration of our model.
We find that ReLU activation function works better with our dataset.
Including kernel regularizers in the NN nodes also reduces the rate of convergence of the model.
The training and validation losses for two NN models are shown in Figure~\ref{fig:loss}, both models have ReLU activation function, without kernel regularizers.
We find that using MSE to compute the error leads to a faster convergence with significant less loss compared to MAE.
The fluctuations in the losses with epochs is also less when using MSE in contrast to noticeable fluctuations when using MAE.
This phenomenon can be explained based on the diversity of our input dataset.
When using MSE as the loss function during training, the penalty imposed on $I$ is higher, which leads to sharp changes in the weights of NN nodes in the layers.
The performance of the two types of models using MSE and MAE against the test data is shown in Figure~\ref{fig:agreement}.
The prediction accuracy for \ce{Kr}, significantly deviate from the agreement line when using MAE metric for the loss function.
The prediction accuracy for the other gases are similar using both MSE and MAE metrics.
We thus focus our work using MSE as the preferred loss function.

\begin{figure}[h!]
    \centering
    \includegraphics[width=0.65\linewidth]{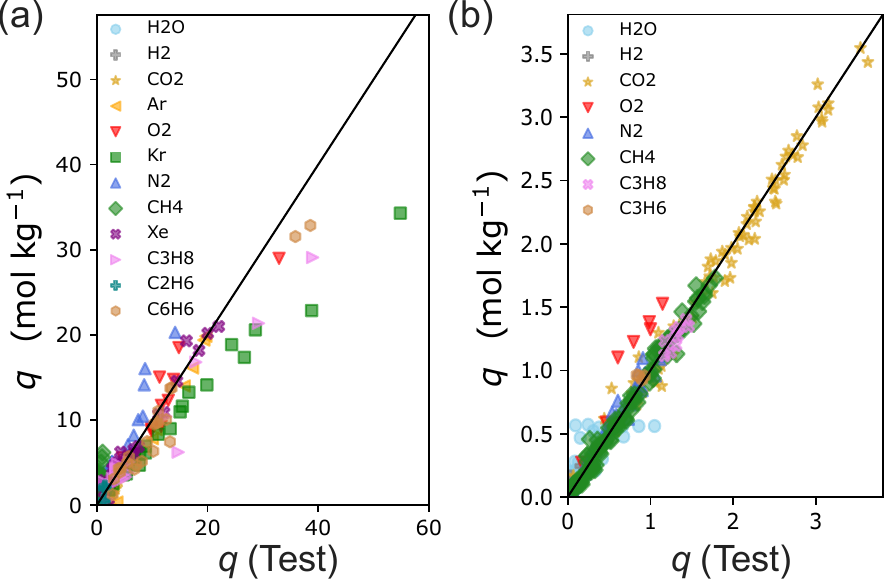}
    \caption{Comparison of gas loading (q) predictions using (a) NN with $(n, h) = (2^{14}, 14)$ trained on the combined dataset and (b) NN with $(n, h) = (2^{10}, 10)$ trained only using $6\text{F-CMSM}$ data. 
    }
    \label{fig:CMS_only_comparison}
\end{figure}

\begin{figure}[h!]
    \centering
    \includegraphics[width=0.7\linewidth]{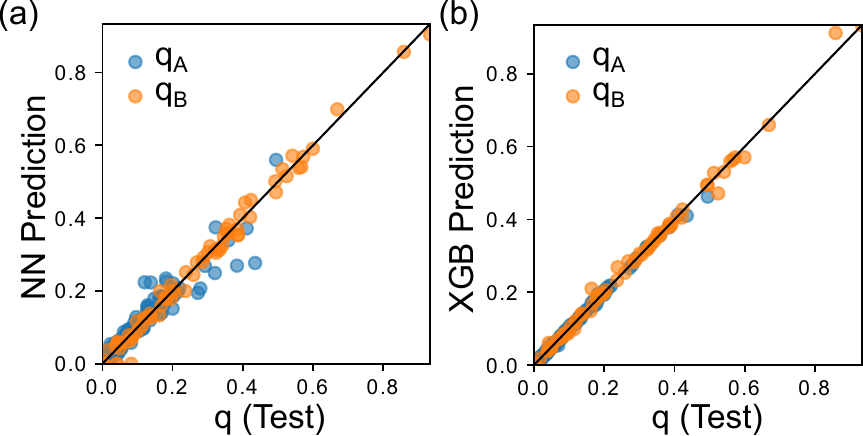}
    \caption{Comparison of gas loading (q) predictions (normalized) using (a) best performing NN having $(n, h) = (240, 3)$ and (b) XGBoost after hyper-parameter optimization.
    The predictions of $q_A$ are for gas loading inside $6$F-CMSM and MOFs.
    $q_B$ denotes the gas loading in cases of binary gas mixtures inside $6$F-CMSM.
    }
    \label{fig:NN_XGB_comparison}
\end{figure}
\begin{figure}[h!]
    \centering
    \includegraphics[width=0.4\linewidth]{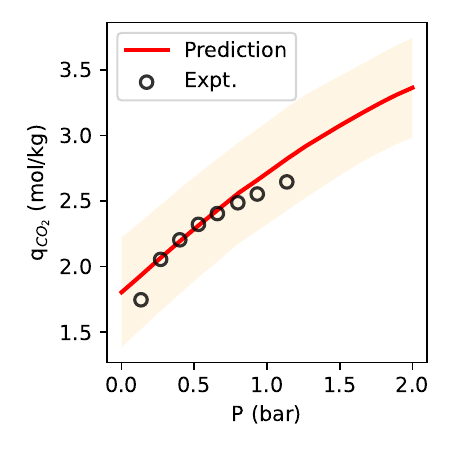}
    \caption{Prediction of \ce{CO2} uptake inside structured CALF-20 framework at $T = \SI{313}{\kelvin}$.
    The red line corresponds to the average prediction obtained from three different neural networks and the shaded region corresponds to its standard deviation.
    The predictions are compared against experimental data points available in literature~\cite{lin2021scalable}.
    }
    \label{fig:calf20_comparison}
\end{figure}

When presented with a given dataset, there is no robust mechanism for deciding what kind of ML architecture one would need to use.
As such, even within an NN model, one cannot ascertain the architecture of nodes and hidden layers \textit{a priori}.
We highlight the importance of choosing a reasonable ML model architecture by studying a set of DNNs having increasingly large $n$.
DNNs require significant computational resources during training and prediction, which becomes their major drawback, and is thus important to use early-stopping on the epochs, once the loss function converges.
By training various DNNs, we observe that the models fail to perform well when trained on the combined polymer and MOF dataset (Figure~\ref{fig:CMS_only_comparison}a).
In contrast, when presented with gas adsorption data inside $6$F-CMSM only (Figure~\ref{fig:CMS_only_comparison}b), the predictions from DNN are close to the ground truth.
This shows that combining two datasets without considering their architectures will lead unsatisfactory predictions.

We perform a thorough search of hyper-parameters for NN and XGBoost models and obtain the best performing ML models as described in the previous section.
Among the trained neural networks, we observe that $(n, h)$ greatly affects, the model's performance.
The best performing $(n, h)$ pairs for both pure and binary gas mixtures are $(100, 3)$, $(240, 3)$, $(256, 3)$, $(260, 3)$, $(320, 7)$, $(330, 5)$, and $(560, 7)$.
For all other combinations, the predictive capabilities are restricted only to the single gas loadings.
In Figure~\ref{fig:NN_XGB_comparison}, we compare the performance of a tuned NN model having $(n, h)=$ (240, 3) vs. tuned XGBoost model.
It is to be noted that the dataset contains Gaussian noise and is thus not susceptible to overfitting.
There is excellent agreement of the XGBoost prediction against our test data.
The NN predictions also have good accuracy, lying very close to the 100\% agreement line.
Although the predictions using XGBoost is closer to the test values, the random forest model does not generate smooth adsorption isotherms for an input gas, temperature, and pressure condition.
The predictions from NN models thus resemble physical processes, while also preserving reasonable accuracy.
To test the performance of our model on unknown data, we study the adsorption isotherm of pure \ce{CO2} within the structured CALF-20 framework ~\cite{lin2021scalable}, at $\SI{313}{\kelvin}$ temperature, which was not present in our training dataset.
The input physical properties for the structured CALF-20 are taken to be $\rho = \SI{0.57}{\gram\per\cubic\centi\meter}$~\cite{nguyen2022separation}, GSA $=\SI{528}{\meter\squared\per\gram}$~\cite{lin2021scalable}, $\xi = 0.35$, $D_i = 2.8 \AA$, and $d_{max} = 4.3 \AA$~\cite{borzehandani2023exploring}.
It is to be noted that CALF-20 shows a drastic increase in \ce{CO2} adsorption in the low pressure regime, unlike the CMS membrane or majority of the MOFs available in our training data.
Despite the lack of such isotherms in training dataset, the models could learn the details of the gas adsorption and could reproduce the uptake capacity with good accuracy.
The obtained isotherm is plotted in Figure~\ref{fig:calf20_comparison} compared alongside with experimental data available in literature~\cite{lin2021scalable}.

\section{Conclusion}
Obtaining precise gas loading capacities by an adsorbent from gas mixtures is important, since a slight deviation can drastically change their solubility coefficients, changing its permeability, affecting the materials expected separation performances.
In this work we predict using machine learning techniques, the gas adsorption isotherms for pure and binary mixtures inside polymers and MOFs.
We utilize only the physical properties of the system to train the ML models on a relatively small dataset, obtained from our simulations and literature.
The ML predictions are in excellent agreement with the available data on $6$F-CMSM and the available subset of the CoRE MOF database.
Moreover, the computational cost associated with training these different ML models is less demanding than performing fully atomistic simulations.
The data sent to the ML model covers a wide range of possible physical values, thereby providing a large range of inputs for the model to predict on, without having to extrapolate.
With available pure gas loadings inside a material, the trained ML models can help map it to a set of input features, which in turn can be used to easily predict the loading capacities in gas mixtures.
Our work thus demonstrates that input features trained only on physical properties of adsorbent materials can predict the gas separation performance of any given class of materials.


\section*{Acknowledgement}

The authors thank the Department of Science and Technology (DST), and Ministry of Education (MOE), India, for providing funding and computational resources.
The authors also acknowledge Science and Engineering Research Board (SERB), India and BRNS-DAE, Govt. of India, for financial support.

\section*{Data availability}
The data used in this work is made available in the Zenodo database \url{https://doi.org/10.5281/zenodo.11567899}


\bibliographystyle{elsarticle-num} 
\bibliography{references}
\end{document}